\documentclass[a4paper,11pt]{amsart}
\usepackage{amssymb}
\begin{document}

\hyphenation{gra-vi-ta-tio-nal re-la-ti-vi-ty Gaus-sian
re-fe-ren-ce re-la-ti-ve gra-vi-ta-tion Schwarz-schild
ac-cor-dingly gra-vi-ta-tio-nal-ly re-la-ti-vi-stic pro-du-cing
de-ri-va-ti-ve ge-ne-ral ex-pli-citly des-cri-bed ma-the-ma-ti-cal
de-si-gnan-do-si coe-ren-za pro-blem gra-vi-ta-ting geo-de-sic
per-ga-mon cos-mo-lo-gi-cal gra-vity cor-res-pon-ding
de-fi-ni-tion phy-si-ka-li-schen ma-the-ma-ti-sches ge-ra-de
Sze-keres con-si-de-red tra-vel-ling ma-ni-fold re-fe-ren-ces
geo-me-tri-cal in-su-pe-rable sup-po-sedly at-tri-bu-table}

\title[Protons do not exert any Hilbertian gravitational repulsion]
{{\bf Protons do not exert any Hilbertian gravitational
repulsion}}

\author[Angelo Loinger]{Angelo Loinger}
\address{A.L. -- Dipartimento di Fisica, Universit\`a di Milano, Via
Celoria, 16 - 20133 Milano (Italy)}
\author[Tiziana Marsico]{Tiziana Marsico}
\address{T.M. -- Liceo Classico ``G. Berchet'', Via della Commenda, 26 - 20122 Milano (Italy)}
\email{angelo.loinger@mi.infn.it} \email{martiz64@libero.it}

\vskip0.50cm

\begin{abstract}
The Hilbertian gravitational repulsion is quite absent in the
Einsteinian field of a proton, owing to the gravitational action
of its electric charge. Accordingly, the proton bunches of the
\emph{LHC} cannot exert any repulsive gravitational force.
\end{abstract}

\maketitle


\normalsize

\vskip1.20cm \noindent \textbf{1.} -- Recently, some authors have
asserted that it would be possible to verify the existence of the
Hilbertian gravitational repulsion by means of the accelerated
protons of the \emph{Large Hadron Collider}. We prove in this Note
that such conviction is erroneous. And for a strong reason.

\vskip1.20cm \noindent \textbf{2.} -- As it is well known, the
Rei\ss ner-Weyl-Nordstr\"om metric of the gravitational field
created by an electrically charged point-mass at rest is (see
\emph{e.g.} our paper ``Attraction and repulsion in spacetime of
an electrically charged mass-point'' \cite{1}, and references
therein):

\begin{equation} \label{eq:one}
\textrm{d}s^{2} = \gamma \, \textrm{d}t^{2} - \gamma^{-1}
\textrm{d}r^{2}
-r^{2}\,(\textrm{d}\vartheta^{2}+\sin^{2}\vartheta\,
\textrm{d}\varphi^{2}) \quad; \quad (G=c=1) \quad,
\end{equation}

\vskip0.20cm where: $\gamma\equiv 1-(2m/r)+(q^{2}/ \,r^{2})$;
$q^{2}\equiv 4\pi \varepsilon^{2}$, and $4\pi \varepsilon$ is the
electric charge of the gravitating point-mass $m$. We have:
$\gamma \,(r=0)=+\infty$; $\gamma \,(r=\infty)=1$; if $q=0$, eq.
(\ref{eq:one}) gives the metric of Schwarzschild manifold in the
standard (Hilbert, Droste, Weyl) form.

\par Two cases: $m^{2}<q^{2}$, in particular $m=0$; and $m^{2}\geq
q^{2}$. \emph{For the electron and the proton} $m^{2}<q^{2}$.

\par We have proved in \cite{1} that \emph{only for} $m^{2}\geq
q^{2}$ there exist spatial regions in which the gravitational
force acts repulsively, both for the radial and the circular
geodesics.

\par If $m^{2}<q^{2}$, $\gamma(r)$ is everywhere positive, with a
minimum value at $r=q^{2} / m$;
$\gamma_{\textrm{min}}=1-(m^{2}/q^{2})$.

\par Quite generally, for the radial geodesics we get the
following first integral:

\begin{equation} \label{eq:two}
\dot{r}^{2} = \gamma^{2}\, (1-|A|\,\gamma)\quad ,
\end{equation}

where: $A<0$ for the test-particles, and $A=0$ for the light-rays.
We have from eq. (\ref{eq:two}) that

\begin{equation} \label{eq:three}
\pm \, \ddot{r} = \gamma \, r^{-2} \, (m-q^{2}\, r^{-1}) \,
(2-3\,|A|\,\gamma) \quad ;
\end{equation}

when $m^{2}<q^{2}$, there is no gravitational repulsion, as it is
easy to prove \cite{1}. An analogous conclusion holds for the
circular geodesics.

\vskip1.20cm \noindent \textbf{3.} -- Now, let us consider a
\emph{distant inertial observer} $\mathrm{\Omega}$, who sees in
motion -- with a given velocity -- a proton which is at rest in
the reference system of sect. \textbf{2}.

\par The inequality $m^{2}<q^{2}$ cannot become $m^{2}>q^{2}$, or $m^{2}=q^{2}$, by
virtue of the motion, in the transformed $\mathrm{\Omega}$-metric
-- and therefore the Lorentzian observer $\mathrm{\Omega}$ does
not register any Hilbertian repulsion exerted by the proton.

\vskip1.20cm \noindent \textbf{4.} -- Some authors believe that in
the \emph{linear} approximation of GR it is possible to compute
directly the gravitational field of a particle of a ``cloud of
dust'', which is in motion with any whatever velocity and
acceleration. Unfortunately, this belief is wrong, because in the
linear approximation -- as in the exact GR -- the equations of
matter motion are \emph{prescribed} by the Einsteinian field
equations, as we have recently emphasized in regard to the
Lense-Thirring effect \cite{2}. Now, for a ``cloud of dust'' of
\emph{neutral} particles, whose energy tensor is $T^{jk}=\mu \,
u^{j} u^{k}$, $(j,k=0,1,2,3)$ -- $\mu$ is the rest density of
mass, and $u^{j}$ the four-velocity -- we have \cite{2}:

\begin{equation} \label{eq:four}
\frac{\partial \, (\mu u^{j})}{\partial \, x^{j}} = 0 \quad, \quad
\textrm{and} \quad \frac{\textrm{d}u^{j}}{\textrm{d}s} = 0 \quad;
\end{equation}

this means that the particle motions are endlessly
\emph{rectilinear and uniform}. The ``dust'' particles do not
interact, they do not create any gravitational field (\emph{in
the} \textbf{\emph{linear}} \emph{approximation}!). For a ``cloud
of dust'' of \emph{electrically charged} particles, whose energy
tensor is again  $T^{jk}=\mu \, u^{j} u^{k}$, if $\sigma$ is the
rest density of charge and $f_{jk}$ the electromagnetic field, one
finds that \cite{3}:

\begin{equation} \label{eq:five}
\frac{\partial \, (\sigma u^{j})}{\partial \, x^{j}} = 0 \quad;
\quad \frac{\partial \, (\mu u^{j})}{\partial \, x^{j}} = 0 \quad;
\quad \mu \, \frac{\textrm{d}u^{j}}{\textrm{d}s} = \sigma \, f^{j}
\: \! _{k} \, u^{k} \quad,
\end{equation}

as a consequence of Maxwell equations and Einstein linearized
field equations. We see that even in this case no gravitational
field is created by the particles of the ``dust''.

\vskip1.20cm \noindent \textbf{5.} -- As a conclusion, we wish to
emphasize the conceptually interesting fact that the very small
gravitational action generated by the electrostatic field of the
proton (and of the electron) is sufficient to suppress any region
of gravitational repulsion \cite{4}.

\end{document}